\documentstyle[twoside,fleqn,espcrc2,epsf]{article}
\newcommand{\be}{\begin{equation}}
\newcommand{\ee}{\end{equation}}
\newcommand{\ba}{\begin{eqnarray}}
\newcommand{\ea}{\end{eqnarray}}

\def\hal{{1\over 2}}
\def\bx{{\bf x}}
\title{Propagators of hot SU(2) gauge theory from 3d adjoint Higgs
 model
 \thanks{Talk
 presented by P. \ Petreczky.
 }}
\author{F. Karsch \address{Fakult\"at
        f\"ur Physik, Universit\"at Bielefeld, D-33615 Bielefeld,
        GERMANY} and P. Petreczky \address{Department for Atomic
        Physics, E\"otv\"os University, P.f. 32, 1518, Budapest,
        HUNGARY}
}       
\begin{document}

\begin{abstract}
We study propagators of the lattice 3d adjoint Higgs model, considered as an
effective theory of 4d $SU(2)$ gauge theory at high temperature. The
propagators are calculated in so-called $\lambda$-gauges. 
From the long distance behaviour of the propagators we extract
the screening masses. It is shown that the pole
masses extracted from the propagators  agree 
well with the screening masses obtained recently in finite
temperature $SU(2)$ theory. The gauge dependence of the screening
masses is also discussed.
\end{abstract}

\maketitle


\section{INTRODUCTION}
The screening of static chromo-electric fields is  one of the most outstanding
properties of $QCD$ at finite temperature 
and its investigation is important both from
a theoretical and phenomenological point of view (for phenomenological
applications see e.g. \cite{wang}).
In leading order of perturbation
theory the associated inverse screening length (Debye mass) is defined
as the $IR$ limit of the longitudinal part of the gluon self energy
$\Pi (k_0=0,{\bf k} \rightarrow 0)$. However, as the screening
phenomenon is related to the long distance behaviour of $QCD$ 
the naive perturbative definition of the Debye mass
is obstructed by severe $IR$ divergences of
thermal field theory and beyond leading order the above definition
is no longer applicable.
Rebhan has shown that the definition of 
the Debye mass through the pole of the longitudinal part of the gluon
propagator is gauge invariant \cite{rebhan}. 
However, this definition requires the introduction of a so-called
magnetic screening mass, a concept introduced long ago \cite{linde}
to cure the $IR$ singularities of finite temperature non-Abelian theories.
Analogously to the electric (Debye) mass the magnetic mass can be
defined as a pole of the transverse part of the finite temperature 
gluon propagator. Because of the IR sensitivity a non-perturbative
determination of the screening mass is necessary.

The main question which we will try to clarify in this contribution is
whether the screening masses,
defined as  poles of the corresponding lattice propagators in 
Landau gauge can be determined from the effective theory (Section 2).
We will consider the simplest
case of the $SU(2)$ gauge group, where precise 4d data on  screening
masses  exist for a huge temperature range \cite{heller}.
We will also discuss the question of gauge dependence of the screening
masses (Section 3).

\section{SCREENING MASSES FROM 3D ADJOINT HIGGS MODEL}
The lattice action for the 3d adjoint Higgs model used in the present paper
has the form
\ba
&&
S=\beta \sum_P \hal Tr U_P + \nonumber\\
&&
\beta \sum_{\bx,\hat i} \hal Tr A_0(\bx) U_i(\bx) A_0(\bx+\hat i)
U_i^{\dagger}(\bx) + \nonumber\\
&&
\sum_{\bx} \biggl[-\beta\left(3+\hal h\right) \hal Tr A_0^2(\bx)
+\nonumber\\ 
&&
\beta x { \left( \hal Tr
A_0^2(\bx)\right)}^2 \biggr],
\label{act}
\ea
where $U_P$ is the plaquette, $U_i$ are the usual link variables and
the adjoint Higgs field is parameterized 
by anti-hermitian matrices $A_0=i \sum_a \sigma^a
A_0^a$ ($\sigma^a$ {are the usual Pauli matrices}). Furthermore 
 $\beta$ is the lattice gauge coupling, $x$ parametrizes the 
quartic self coupling of the Higgs field and $h$ denotes the
bare Higgs mass squared. This model is known to have two phases: the
broken (Higgs) phase and the symmetric (confinement) phase separated by
a line of $1^{st}$ order phase transitions \cite{hart,kajantie1}. 
The high temperature phase 
of the 4d $SU(2)$ gauge theory corresponds to some surface in the
parameter space $(\beta, h, x)$, the surface of 4d physics
$h=h_{4d}(x,\beta)$. This surface may lie in the symmetric phase or in
the broken phase, i.e. the physical phase might be either the symmetric
or the broken phase. The surface of 4d physics determined by the
2-loop level dimensional reduction \cite{kajantie1} lies in the broken
phase. This fact appears to be self-contradictory because
if $g\ll 1$ for expectation value of $A_0$ one has
$A_0 \sim 1/g$ but the dimensional
reduction is only valid if $A_0 \ll \pi T$.
To overcome this difficulty of the dimensional reduction the authors of
Ref. \cite{kajantie1} suggested to use the $1^{st}$ order nature of the 
phase transition and perform measurements at the values of the
parameters obtained from the 2-loop dimensional reduction but in
the metastable phase \footnote{In a finite volume there always is a
region in the parameter space where the symmetric and broken phases are 
metastable. Only in the infinite volume limit an assignement to one of
these phases will become possible.  
In fact, on lattices typically used in numerical calculations 
the surface of 4d physics lies in this metastable region
\cite{kajantie1}.}.
The obvious problem with this approach is that the metastable phase
disappears in the infinite volume limit.
Therefore in Ref. \cite{ours} simulations were done in the symmetric phase
and it was suggested to determine the surface of 4d physics by
non-perturbative matching.

In this section we are going to review our results on electric and
magnetic screening masses obtained from the Landau gauge propagators in
the symmetric phase as well as in the metastable 
region of our finite lattices. 
Most of our numerical studies have been performed on lattices of size
$32^2\times 64$ and at $\beta=16$. 
Simulations in the metastable region have been performed at the
values of the parameters obtained from 2-loop dimensional reduction 
\cite{kajantie1}. 
The two sets of values of $h$ and $x$ used in our simulations in the
symmetric phase are shown in Table 1, where also the corresponding
temperature values as well as the values of $h$ corresponding to the 
2-loop dimensional reduction ($h_{4d}$) are indicated. The temperature scale is	
essentially fixed by $x$. The detailed procedure of choosing the
parameters in the symmetric phase is given in Ref. \cite{ours}.
\begin{figure}
\epsfysize=6cm
\epsfxsize=8cm
\centerline{\epsffile{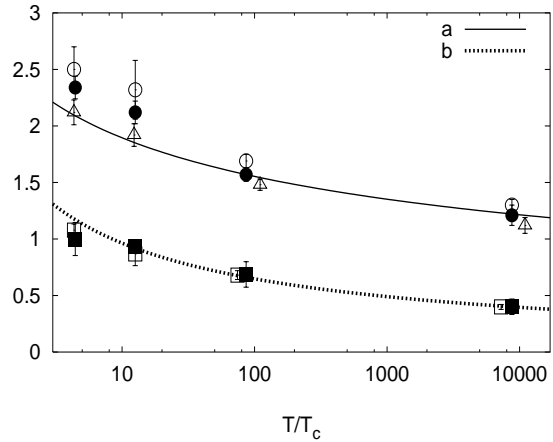}}
\vspace{-0.7cm}
\caption{The screening masses in units of the temperature. 
Shown are the Debye mass
$m_D$ for the first (filled circles) and the second (open circles)
set of $h$, and
the magnetic mass  $m_T$ for the first (filled squares)
and the second (open squares) set of $h$.
The line (a) and line (b) represent the fit for the temperature
dependence of the Debye and the magnetic mass from 4d
simulations
The open triangles are the values of the Debye mass for $h_{4d}(x,\beta)$
obtained from the 2-loop dimensional reduction in the metastable region.
Some data points at the temperature $T \sim 90 T_c$
and $T \sim 9000 T_c$ have been shifted in the temperature scale for
better visibility.}
\vspace{-0.5cm}
\end{figure}
The results on the screening masses are summarized in
Figure 1 where also the results of 4d simulations \cite{heller} are
shown. \\
\vskip0.2truecm
\noindent Table 1\\ 
The two sets of the
bare mass squared used in the simulation and those
which correspond to the 2-loop dimensional reduction for $\beta=16$
\begin{flushleft}
\begin{tabular}{|l|}
\hline 
$~~x~~$  $~~T/T_c~~$~~~~~$~~~~h_I~~~~$~~~$~~~~h_{II}~~$~~$~~h_{4d}~~$\\
\hline
$~0.09~$  $~4.433$~   $~-0.2652$   $~-0.2622$  $~-0.2700$\\
$~0.07~$  $~12.57$~   $~-0.2528$   $~-0.2490$  $~-0.2588$\\
$~0.05~$  $~86.36$~   $~-0.2365$   $~-0.2314$  $~-0.2437$\\
$~0.03~$  $~8761$ ~   $~-0.2085$   $~-0.2006$  $~-0.2181$\\
\hline
\end{tabular}
\end{flushleft}

As one can see from the figure the agreement between the masses
obtained from 4d and 3d simulation is rather good. The magnetic mass
practically shows no dependence on $h$ and its value is rather close to
the magnetic mass of 3d pure gauge theory $m_T=0.46(3)g_3^2$
\cite{ours} ($g_3^2$ is
the 3d gauge coupling). The electric mass shows
some dependence on $h$. For relatively low temperatures ($T<50 T_c$)
the best agreement with the 4d data for the Debye mass 
is obtained for values of $h$
corresponding to 2-loop dimensional reduction and lying in the
metastable region. For higher temperatures, however, practically no distinction
can be made between the three choices of $h$. 

\section{GAUGE DEPENDENCE OF THE SCREENING MASSES}
Let us turn to the discussion of the gauge dependence of the propagator
masses. 
The propagator pole mass was proven to be gauge independent
at any given order of perturbation theory \cite{kobes}.
Whether this holds also non-perturbatively is, however, an open
question \footnote{The propagator pole in the 3d adjoint Higgs model does
not correspond to an asymptotic state ( the theory is confining) 
and therefore there there is a priori no reason for gauge independence 
of the pole mass. 
}.
To study the gauge dependence of the pole masses we have used
the so-called $\lambda$-gauges \cite{bernard} defined by the gauge 
fixing condition 
\be
\lambda \partial_3 A_3+\partial_2 A_2+\partial_1 A_1=0.
\ee
The case $\lambda=1$ corresponds to the Landau gauge. 
For the numerical analysis the following values of the parameter
$\lambda$ have been chosen: $\lambda=0.5,~1,~2,~8$.
We have measured the electric and the magnetic correlators on
a $32^2\times 96$ lattice at $x=0.03$ and two values of $\beta$ and $h$:
$\beta=16,~h=-0.2085$ and $\beta=24,~h=-0.1510$ 
both corresponding to the symmetric phase.
The results of these measurements for the electric ($A_0$) propagator
are shown in Figure 2. 
As one can see from the Figure the large distance behaviour of 
the propagators seems to exhibit some gauge dependence 
which becomes visible for large values of $\lambda$. 
The situation is similar for the magnetic propagator \cite{petrphd}.  
To what extent
this behaviour is influenced by the gauge fixing procedure used by us
requires further analysis. Here 
one also should consider
gauges other than $\lambda$-gauges \cite{cucc}.
\begin{figure}
\epsfysize=6cm
\epsfxsize=8cm
\centerline{\epsffile{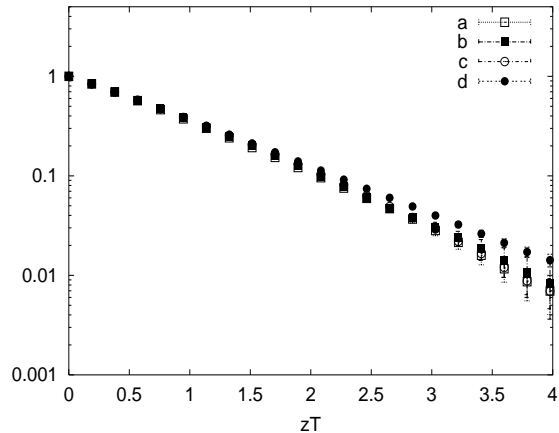}}
\vspace{-0.7cm}
\caption{ The electric ($A_0$) propagators at
$x=0.03,~h=-0.1510,~\beta=24$ for different values of $\lambda$:
$\lambda=0.5$ (a), $1.0$ (b), $2.0$ (c), $8.0$ (d), 
}
\vspace{-0.5cm}
\end{figure}

{\bf Acknowledgments:} This work was partly supported by the TMR 
network ERBFMRX-CT97-0122 and the DFG under grant Ka 1198/4-1. 
The numerical work has been performed
at the HLRZ J\"ulich and the HLRS Stuttgart.

\end{document}